# In-depth Understanding of the Band Alignment and Interface States Scenario in $Bi_2O_2Se$/$SrTiO_3$ Ultrathin Heterojunction

Ke Zhang[1,*], Yusen Feng[1], Lei Hao[2,3*], Jing Mi[2,3], Miao Du[2,3], Minghui Xu[1], Yan Zhao[1], Jianping Meng[4], and Liang Qiao[1,*]

Affiliations

[1)] School of Physics, University of Electronic Science and Technology of China, Chengdu 610054, China

[2] GRINM Group Co., Ltd., National Engineering Research Center of Nonferrous Metals Materials and Products for New Energy, Beijing 100088, China

[3] GRIMAT Engineering Institute Co., Ltd., Beijing 101407, China

[4)] Beijing Institute of Nanoenergy and Nanosystems, Chinese Academy of Sciences, Beijing 101400, China

[*] Authors to whom correspondence should be addressed: phyzk@uestc.edu.cn (K. Zhang), haolei863@163.com (L. Hao). and liang.qiao@uestc.edu.cn (L. Qiao)

K. Zhang and Y. Feng contributed equally to this work.

**Abstract** Bismuth oxyselenide ($Bi_2O_2Se$), a novel quasi-2D charge-carrying semiconductor, is hailed as one of the best emerging platforms for the next generation semiconductor devices. Recent efforts on developing diverse $Bi_2O_2Se$ heterojunctions have produced extensive potential applications in electronics and optoelectronics. In-depth understanding of the band alignment and especially interface dynamics is, however, still challenging. In this work, a comprehensive experimental investigation on the band alignment is performed by a high-resolution X-ray photoelectron spectrometer (HRXPS), and the properties of interface states are also fully discussed. The results show that the ultrathin film $Bi_2O_2Se$ grown on $SrTiO_3$ ($TiO_2$ (001) termination) exhibits Type-I (straddling gap) band alignment with a valence band offset (VBO) of about 1.77±0.04 eV and conduction band offset (CBO) of about 0.68±0.04 eV. However, further considering the contribution of the interface states, the bands on the interface present a herringbone configuration due to sizable build-in electric fields, which is significantly different from the conventional band alignment. In this sense, our results provide an insightful guidance to the development of high-efficiency electronic and optoelectronic devices, specifically of the devices where the charge transfer is highly sensitive to interface states.

## 1 Introduction

Owing to ultrahigh mobility (@1.9K, ~29000$cm^2V^{-1} \cdot s^{-1}$; @room temperature, 450$cm^2V^{-1} \cdot s^{-1}$) [1,2], outstanding air stability[3], moderate and tunable bandgap (~0.8eV and thickness dependent) [2,4], as well as the virtue of quick response to various input signals [5,6], $Bi_2O_2Se$ is expected as an emerging material platform for the next-generation electronic industry [7–11]. Recently, diverse applications of $Bi_2O_2Se$ are being extensively studied, including optoelectronic devices [12], field-effect transistors [2], memristors [6], photodetector [13,14], true random number generator and sensors [15–17]. Among them, massive efforts on combining 2D $Bi_2O_2Se$ with other materials, such as graphene [18,19], transition metal dichalcogenides (TMDs) [20–23], topological materials [24] etc., have produced many sophisticated artificial heterojunctions with novel properties.

Nevertheless, as is well known, "the interface is the device" [25] and "in the 2D limit, the properties of the interface govern everything" [26]. The band offsets and charge transfer properties have been examined in many papers so far [5,20,23,27–29]. Nevertheless, these work underestimate the role of the interface states. Thus, the true band alignment, particularly, in-depth understanding of the electrons and holes transport properties, the build-in potential strength, and the contribution of interface states remain elusive, despite its importance in harnessing their preeminent functionalities.

In this work, by pulsed laser deposition (PLD) epitaxial growth method, we have precisely fabricated atomic-scale flat heterojunctions formed by growing $Bi_2O_2Se$ on $SrTiO_3$ ($TiO_2$ (001) termination). HRXPS was carried out to determine the band alignment of the $Bi_2O_2Se/SrTiO_3$ heterojunction (termed as BOS/STO) by extrapolating the VBO and the CBO at the interface. Our work reveals that the BOS/STO heterojunction exhibits Type-I band alignment with VBO of about 1.77±0.04 eV and CBO of about 0.68±0.04 eV. Considering the contribution of the interface states and *pd* orbital coupling, the bands on both sides of the BOS/STO interface prefer a herringbone bending configuration with an extremely large build-in electric field ($E_{build-in}^{BOS}$) estimated greater than 10mVÅ$^{-1}$. Such a large electric field breaks the local inversion symmetry, lowering the bulk symmetry form centrosymmetric $D_{4h}$ to polar $C_{4v}$ symmetry. Further, combining the intrinsic strong spin-orbital-coupling (SOC), hence, a giant Rashba-type spin splitting in the heterojunction region is introduced [30]. A virtue of this coupling is that it opens up the possibility of manipulating the spin degree of freedom of carriers without external magnetic field [31]. Thus, this work provides insight for in-depth understanding the band bending and the properties of interface states in such 2D heterostructures, and also sheds light on the feasibility of integrating 2D $Bi_2O_2Se$ with other materials to fabricate heterostructures with efficient charge and spin transfer.

## 2 Results and discussion

The band alignment was judged by XPS method which has been extensively utilized to measure the VBO of heterojunctions [32–39]. The valence band maximum (VBM) and core level (CL) binding energies were measured by our HRXPS instrument (*SPECS*, Phoibos150) with a monochromated X-ray source (Al Kα, $h\nu$=1486.61 eV). High-resolution narrow and survey spectra were collected at fixed analyzer transmission mode with energy resolutions of 10 meV and 50 meV, respectively. During the measurement, a low-energy electron flood gun was required for charge neutralization. Binding energies were calibrated by the $C_{1s}$ peak of adventitious carbon, which is taken to be 284.8 eV [33,40]. The peak deconvolution of all the elements was carefully accomplished with well agreement of the SOC splitting strength for double peaks, and then fitted by Voigt function after background deduction. The VBM was determined by extrapolating a linear fit of the leading edge of the valence band to the baseline.

For a specific band alignment, a set of samples, consisting of bulk $SrTiO_3$ substrate (material A), thick (~20nm) BOS/STO heterojunction which was regarded as bulk $Bi_2O_2Se$ (material B) and an ultrathin epilayer (~3nm) BOS/STO heterojunction (HJ_AB), were prepared (Detailed method please refer to supporting information S1). The CL energies of materials A and B along with their VBM energies were obtained from the HRXPS spectra. The HJ_AB provides the difference in the CL energies on the interface of materials A and B. Hence, the VBO ($\Delta E_{VBO}^{AB}$) of heterojunction can then be determined using the following method developed by Kraut et al [41,42].

$$\Delta E_{VBO}^{AB} = (E_{CL}^{A} - E_{VBM}^{A})_{material\ A} - (E_{CL}^{B} - E_{VBM}^{B})_{material\ B} - (E_{CL}^{A} - E_{CL}^{B})_{HJ\_AB} \quad (1)$$

Consequently, the CBO ($\Delta E_{CBO}^{AB}$) can be deduced combining each bandgap ($E_g$) of material A and B by the following equation:

$$\Delta E_{CBO}^{AB} = (E_g)_{material\ A} - (E_g)_{material\ B} - \Delta E_{VBO}^{AB} \quad (2)$$

Fig. 1 shows the high-resolution CLs and survey spectrum of STO substrate (material A). The CLs of titanium and strontium are shown in panel (a)~(c), which are deconvoluted by two peaks according to the SOC splitting strength. For example, in panel (a) a narrow peak locating at 458.59eV and a fat peak locating at 464.22eV correspond to Ti $2p_{3/2}$ and Ti $2p_{1/2}$ split peak respectively, showing strong spin orbital split strength with a magnitude of $\Delta E_{Ti2p}$ =5.63eV. Besides, the full width at half maximum (FWHM) of Ti $2p_{3/2}$ is smaller than that of Ti $2p_{1/2}$, due to the Coster-Kronig effect, agreeing well with other reports [43]. Panel (b) and Panel (c) depict the $3p$ and $3d$ orbitals of strontium, where the binding energies of 268.62eV and 133.12eV correspond to main peaks of Sr $3p_{3/2}$ and Sr $3d_{5/2}$, respectively. Panel (d) presents the valence band spectrum where the VBM is obtained by linearly extrapolating the lead edge to the baseline of the background intensity [36]. Here the VBM of the STO is measured to be ~ 2.42eV, confirming its *n* type property. Thereby, the difference values between the CLs and VBM [$(E_{CL}^{A} - E_{VBM}^{A})_{material\ A}$] for STO are calculated to be 456.17eV, 266.20eV and 130.70eV for Ti $2p_{3/2}$, Sr $3p_{3/2}$ and Sr $3d_{5/2}$, respectively, as shown in Fig.1.

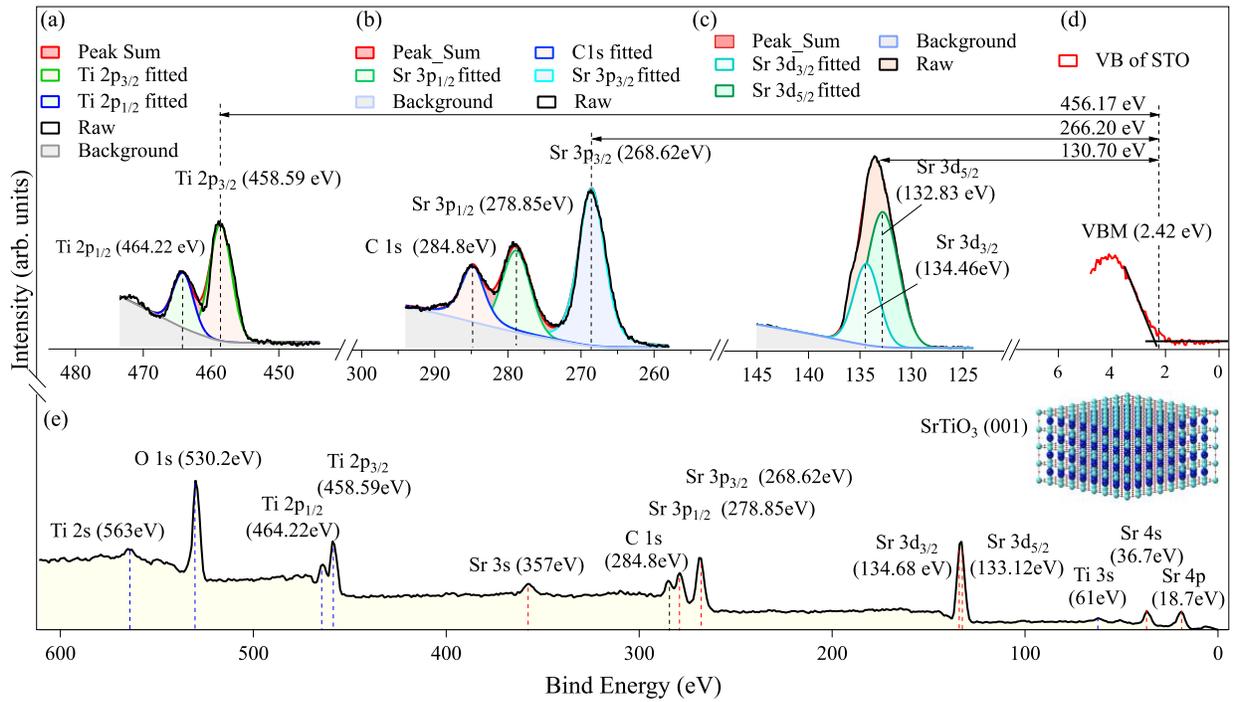

Fig. 1. High resolution X-ray photoelectron spectra (HRXPS) of STO. CLs of (a) Ti $2p$, (b) Sr $3p$, (c) Sr $3d$, (d) VBM and (e) survey spectra.

Similarly, Fig. 2 and Fig. 3 show high-resolution CLs and survey spectra of ~20nm BOS/STO (material B) and ~3nm BOS/STO (HJ_AB), respectively. The measured CLs for all the samples are listed in Table 1. It is easy to know that the difference value of the CL and VBM of the 20nm BOS/STO [$(E_{CL}^{B} - E_{VBM}^{B})_{material\ B}$] is 441.19eV (157.94eV) for the main peak of Bi 4$d_{5/2}$ (Bi 4$f_{7/2}$). In the 3nm BOS/STO sample, the difference values of the CLs $[(E_{CL}^A - E_{CL}^B)_{HJ\_AB}]$ are 16.76eV between Ti 2$p_{3/2}$ and Bi 4$d_{5/2}$, 155.42eV between Bi 4$d_{5/2}$ and Sr 3$p_{3/2}$, 110.02eV between Sr 3$p_{3/2}$ and Bi 4$f_{7/2}$, and 300.2eV between Ti 2$p_{3/2}$ and Bi 4$f_{7/2}$, respectively. Consequently, through equation (1), one can easily calculate the VBOs for different CLs, such as $\Delta E_{VBO}^{Bi4d_{5/2}-Ti2p_{3/2}}$=1.78eV, $\Delta E_{VBO}^{Bi4f_{7/2}-Ti2p_{3/2}}$=1.97eV, $\Delta E_{VBO}^{Bi4d_{5/2}-Sr3p_{3/2}}$=1.57eV, and $\Delta E_{VBO}^{Bi4f_{7/2}-Sr3p_{3/2}}$=1.76eV. Thereby, the resultant average value of the VBOs ($\Delta E_V$) of the 3nm BOS/STO is supposed to be 1.77eV.

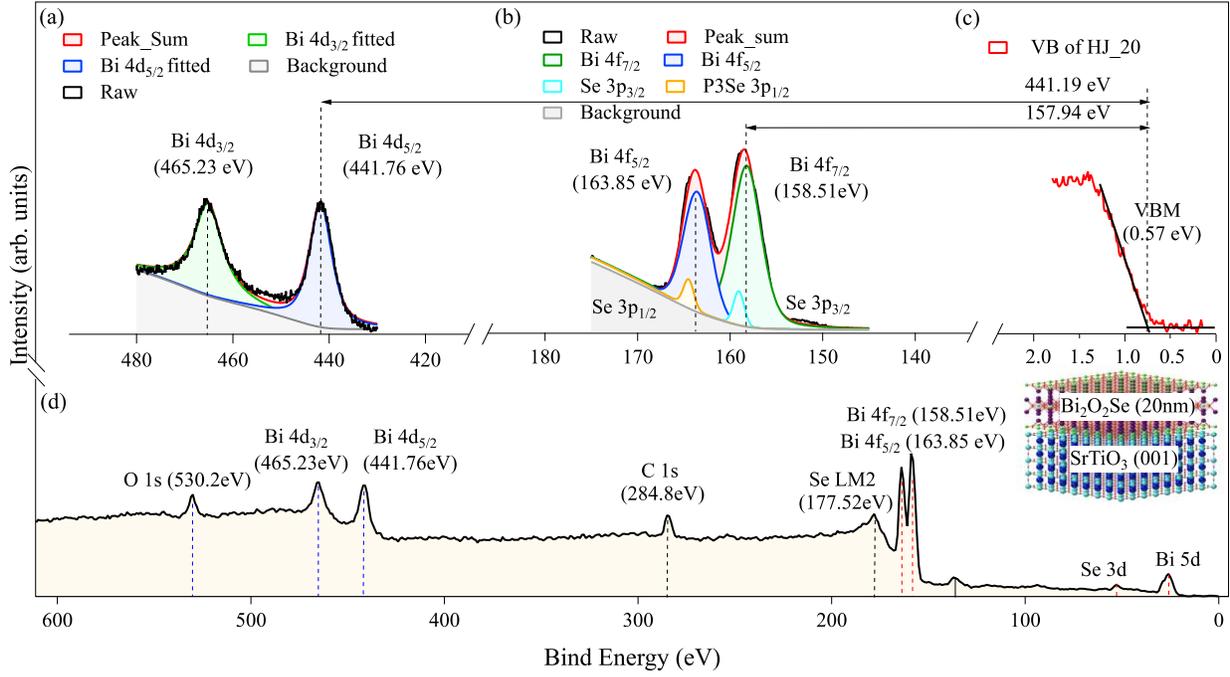

Fig. 2. High resolution X-ray photoelectron spectra (HRXPS) of bulk 20nm BOS/STO. CLs of (a) Bi 4$d$, (b) Bi 4$f$, (c) VBM, and (d) survey spectra.

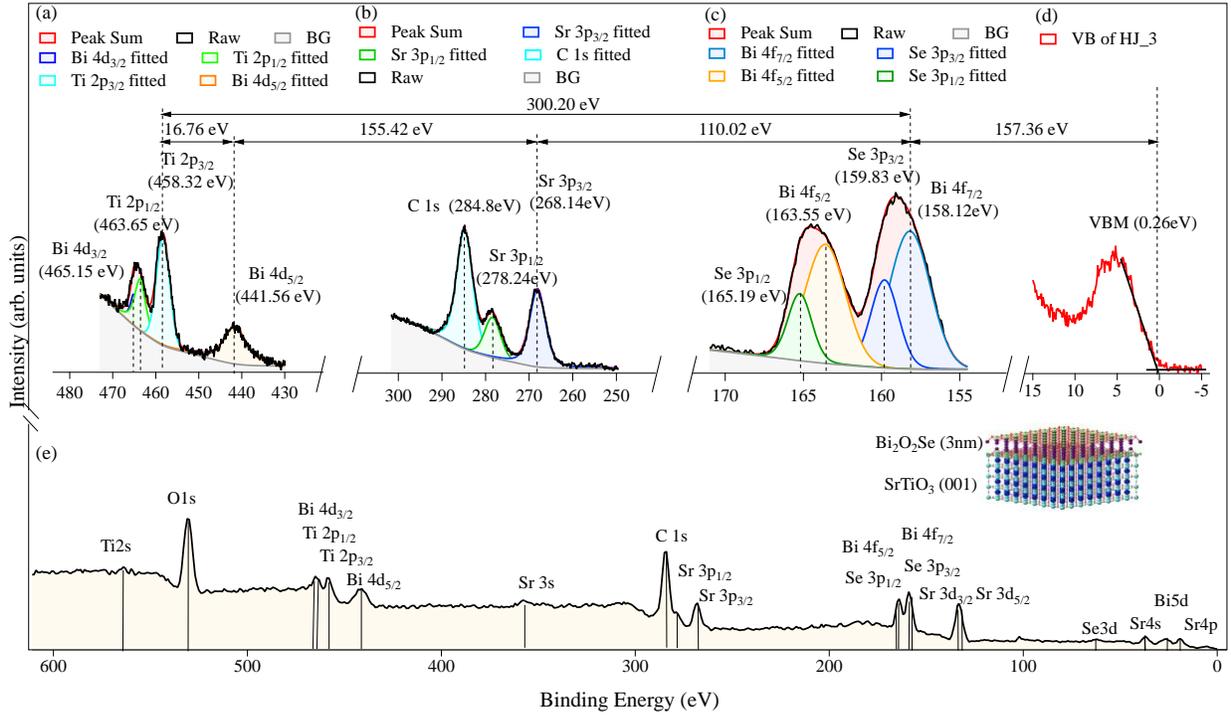

Fig. 3. High resolution X-ray photoelectron spectra (HRXPS) of 3nm BOS/STO. CLs of (a) Ti 2*p*, (b) Sr 3*p*, (c) Se 3*p* and Bi 4*f*, (d) VBM and (e) survey spectra.

Next, in equation (2), by substituting the estimated average VBO ($\Delta E_V$=1.77eV) and the energy gap of each bulk STO (3.25eV [44]) and bulk BOS (0.8eV [2,45]), one can easily quantitate the CBO ($\Delta E_C$) to be 0.68eV. Now, let's take the errors into account. For VBMs, the linear extrapolation method consistently yields correct VBMs with an accuracy of about $\Delta_{VBM}$=~±0.02 eV especially for high resolution spectrum [41]. For CLs fitting, even with great care in dealing the split strength and FWHM of the doublets, there is, inevitably, an uncertainty ($\Delta_{CL}$=~±0.03eV) due to fitting program and empirical arbitrariness (More details on fitting please refer to supporting information, S2). For surface charging, to rule out differential charging, an electron flood gun was utilized for charge compensation. However, the magnitude of any differential charging is expected to be within $\Delta_{Charging}$=~±0.02eV according to *SPECS* user guide. For strain, due to lattice mismatch, there should be some strain in epilayer, depending on specific materials and lattice mismatch. In this case, it generally less than $\Delta_{Strain}$=~±0.01eV [45]. As for the energy gaps of BOS and STO, we adopt the ARPES and UV-Vis optical absorption experiment results, thereby, the gap errors are not considered in this work. Therefore, the resultant total errors of the VBO ($\Delta E_V$) and CBO ($\Delta E_C$) is estimated to be $\Delta_{Total} = \pm\sqrt{\Delta_{VBM}^2 + \Delta_{CL}^2 + \Delta_{Charging}^2 + \Delta_{Strain}^2} = \pm 0.04$eV.

Table I. HRXPS CL energies, VBM energies and estimated average VBO ($\Delta E_V$) and CBO ($\Delta E_C$) of the STO, BOS and 3nm BOS/STO samples.

| | Ti 2$p_{3/2}$(eV) | Bi 4$d_{5/2}$(eV) | Sr 3$p_{3/2}$(eV) | Bi 4$f_{7/2}$(eV) | VBM | $\Delta E_V$ | $E_g$ | $\Delta E_C$ |
|---|---|---|---|---|---|---|---|---|

| | | | | | | |
|---|---|---|---|---|---|---|
| STO bulk | 458.59 | | 268.62 | | 2.42 | 3.25 |
| BOS bulk (20nm BOS/STO) | | 441.76 | | 158.51 | 0.57 | 0.8 |
| 3nm BOS/STO | 458.32 | 441.56 | 268.14 | 158.12 | 0.26 | 1.77 | 0.68 |

The experimentally determined parameters VBO=1.77±0.04eV and CBO=0.68±0.04eV indicate that the BOS/STO heterojunction can be fitted into type-I (straddling) band alignment shown in Fig. 4. In panel (a), one can directly draw a simplified band alignment diagram, just delivering the straddling configuration of type-I. However, prior forming equilibrium heterojunction, the Fermi lever of the BOS is lower than that of STO, so the electrons in conduction bands of STO will transfer to the surface of the BOS. Thus, a build-in electronic field pointing from STO to BOS is formed, which will result in the band of STO side upward bending while the band of BOS side downward bending, as shown in panel (b). Be advised, it is not difficult to find that in this diagram the VBM of the heterojunction (VBM$_{HJ}$) is must larger than that of BOS (labelled as VBM$_1$ in panel (b)). However, this is inconsistent with our XPS experiment result that the VBM$_{HJ}$ (0.26eV) is smaller than VBM$_1$ (0.57eV). This inconsistent will further extend to the conduction bands, which will result in a dilemma that the CBO$_{HJ}$ (ΔE$_C$=0.68eV) should larger than the difference between the conduction bands to the Fermi level (δ$_2$=0.83eV).

To figure out this problem, one must take the contribution of the interface states into account. As well known, there are interface states in the bandgap at the surface of the semiconductor (For experimental evidences on interface or surface states please refer to supporting information, S3). The interface states are generally divided into two types: donor type and acceptor type. For the donor interface states, the energy levels are electrically neutral when occupied by electrons and positively charged when the electrons are discharged. Conversely, for the acceptor interface states, the energy levels are electrically neutral when they are unoccupied and negative when they are electron accepted. For most *n*-type semiconductors, the surface Fermi level in the band gap is located about one-third of the way from the top of the valence band. For example, if the surface levels are filled with electrons, the surface is negatively charged, showing an acceptor type. To be more specific, suppose in an *n*-type semiconductor and there are acceptor surface states (E$_{AF}$) above the Fermi level, then the energy levels between E$_F$ and E$_{AF}$ is basically filled with electrons, scilicet the surface is negatively charged. Thus, positive charges must appear near the surface of the semiconductor on both sides. Consequently, the build-in electric fields are both pointing to the interface, leading to the bands on both sides upward bending, as shown in Fig. 4(c), in contradistinction to the band diagram in Fig. 4(b) where the role of the interface states are not taken into account.

Now, in this scenario, things get to make sense. As shown in Fig. 4(c), the interface states lie in the one-third of the gaps, resulting in the bands upward bending, showing a tip-tilted band diagram, in other words a herringbone configuration. Consequently, according to measured VBO (ΔE$_V$) and CBO (ΔE$_C$), one can easy to obtain the bending strength of both bands. For example, the $qV_1 = VBM_1 - VBM_{HJ} = 0.31 eV$, $qV_2 = VBM_2 - VBM_{HJ} - \Delta E_V = 0.38 eV$, $\delta_1 = E_{g1} - VBM_1 = 0.23 eV$, $\delta_2 = E_{g2} - VBM_2 = 0.83 eV$. Suppose the work function of STO ($W_1$) is 4.58eV [46], then one can directly estimate the electron affinity potential of STO and BOS to be $\chi_2 = 3.75 \ eV$ and $\chi_1 = 4.35 eV$, respectively. To check the correctness of this band diagram, one can check the $\delta_2 = \Delta E_C + qV_1 + \delta_1 - qV_2 = 0.84 eV$, which is within the error of the measurement, exhibiting near-perfect self-consistency.

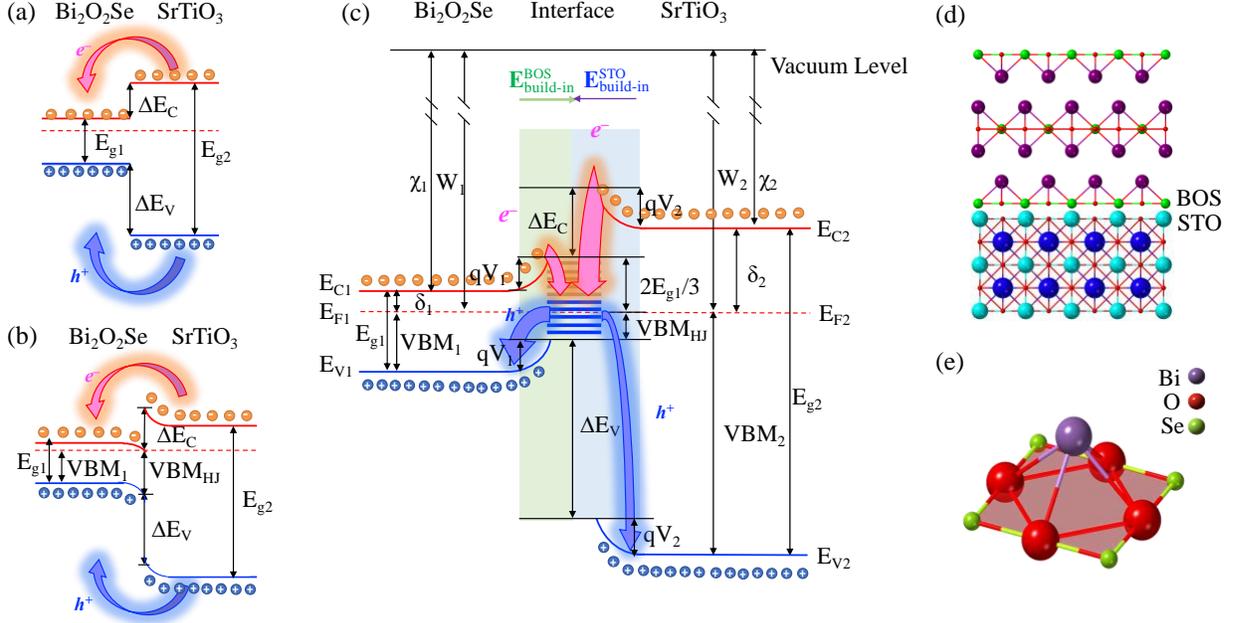

Fig. 4 Band diagrams for BOS/STO band alignment. (a) simplified band diagram, (b) and (c) band diagrams without and with the consideration of the interface states, (d) schematic of crystal lattice mismatch and (e) the bond configuration of (001) plane.

This scenario is fundamentally supported by the relationship between the dangling bonds and the interface states. In heterojunctions, due to inevitable lattice mismatch, dangling bonds extensively exist on the interface, introducing interface states in the bandgap. Fig. 4(d) shows a schematic diagram of the lattice mismatch. As seen, the dangling bonds appear in the smaller lattice side. Further, one can estimate the density of the dangling bonds easily. In the (001) plane of BOS, the area of a unit cell is $a_1^2$, where $a_1 (= 3.887 Å)$ is the lattice constant. The (001) plane contains 4 bonds, as shown in Fig. 4(e). Therefore, the bond density of the crystal plane (001) is $4/a_1^2$. For the heterojunction formed by two semiconductors with lattice constants $a_1$ and $a_2 = 3.905 Å$ (where $a_2$ is lattice constant of STO), the suspension bond density is estimated about $4 \left[\frac{a_2^2 - a_1^2}{a_1^2 a_2^2}\right] \approx 2.4 \times 10^{13} cm^{-2}$. According to Schottky and Bardeen limits, when density of the interface states is greater than or equal to $10^{13} cm^{-2}$, the Fermi level lies in the one-third of the bandgap and the dangling bonds work as acceptors in $n$-type semiconductors.

According to Zunger's common-anion rule [47,48] and $pd$ orbital coupling, one can also understand this band alignment in another approach. For an arbitrary AX/BX heterojunction containing common anion $p$ states, the common-anion rule prefers to have a negligible VBO, as shown in Fig.5 (a). However, cation $d$ orbitals couple with anion $p$ states, resulting in VBM upward shift when cation A m"$d$ (B n"$d$) occupied orbital lies below the anion A-X $p$ (B-X $p$) orbital, while VBM downward shift when cation A m"$d$ (B n"$d$) unoccupied orbital lies above the anion A-X $p$ (B-X $p$) orbital [33]. Hence, there are two coupling types as shown in Fig.5(b) and (c). In the BOS/STO heterojunction, it belongs to the $pd$ coupling-II case, in which the Bi 5$d$ orbitals in BOS lie below the O 2$p$ and Se 4$p$ orbitals, leading to a up shift of the VBM$_1$ of BOS. Whereas in STO, the Ti 3$d$ orbitals lie above the O 2$p$ orbitals, resulting in a downward energy shift of the VBM$_2$ of STO. Therefore, a large VBO in a type-I band alignment is formed, as shown in Fig. 5(d). Next, based on the measured VBO and CBO, now we are able to estimate the build-in electric field. In the BOS epilayer, the depth of the carries transfer is conservatively estimated as long as the thickness ($d_{max} \sim 3$nm). According to the build-in potential ($qV_1 = 0.31 eV$), we thus estimate the minimum of the build-in electric field to be $E_{build-in}^{BOS} \geq qV_1/d_{max} = 10 meV/\text{Å}$. In fact, the carriers transfer only take place on the surface of the BOS, the build-in electric field should be much larger than $10 meV/\text{Å}$. Such a large build-in electric field undoubtedly breaks the inversion symmetry, lowering the intrinsic centrosymmetric $D_{4h}$ symmetry to polar $C_{4v}$ symmetry. Consequently, a large spin polarization is strongly expected without external field in this nonmagnetic BOS/STO system. From a spin device perspective, this system is more realistic compared with monolayer [49] or polar and nonpolar cleavage surface engineering [31].

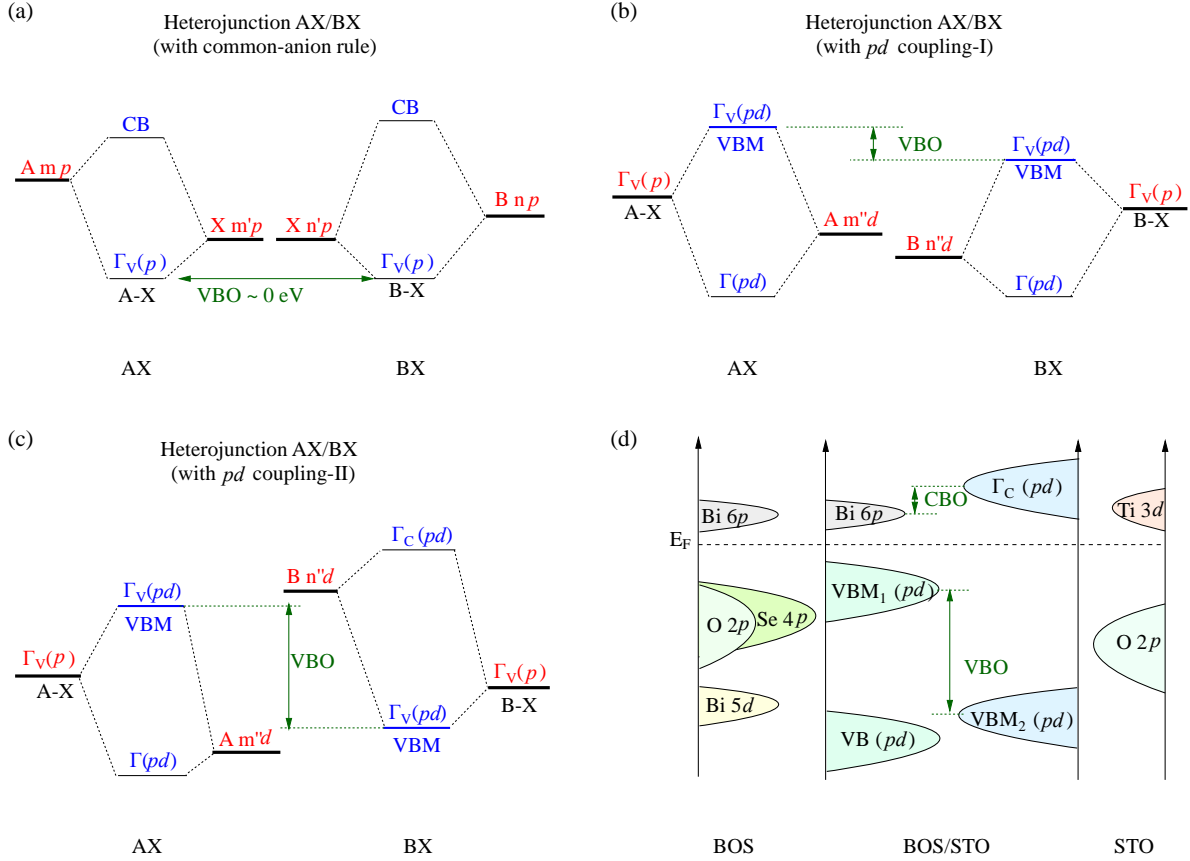

Fig. 5 Schematic drawing of common-anion rule and VBMs shift in the case of BOS/STO. (a) negligible VBO with the absence of *pd* coupling, (b) and (c) *pd* coupling-I and -II, respectively, and (d) common-anion rule and *pd* coupling-II in the case of BOS/STO.

## 3 Conclusion

In this work, we systematically studied the band offset parameters through HRXPS experiments. The BOS/STO heterojunction exhibits type-I band alignment with VBO of 1.77±0.04eV and CBO of 0.68±0.04eV. Consequently, the band diagram with interface states was plotted, in which the bands on both sides of the heterojunction present tip-tilted bending due to the acceptor-type interface states around the Fermi level. And an extremely large build-in electric field ($E_{build-in}^{BOS}$) is estimated great larger than 10mVÅ$^{-1}$. Such scenario provides insight for in-depth understanding the band bending and the properties of interface states in such 2D heterostructures, and also sheds light on the development of high-efficiency electronic and optoelectronic devices, specifically of the devices where the charge transfer is highly sensitive to interface states.

**Acknowledgements** This work is supported by National Natural Science Foundation of China (NSFC) (Grants Nos. 52072059, 12304078, 12274061 and 11774044) and Natural Science Foundation of Sichuan Province (Grants No. 2024NSFSC1384).

**Declarations**

**Conflict of interests** The authors declare that they have no conflict of interest.